\documentstyle[12pt]{article}
\title{Modified Kantowski-Sachs Cosmologies from Coset Models}
\author{Leopoldo A. Pando Zayas\thanks{E-mail address:leopoldo@grg1.phys.msu.su}\\
Department of Theoretical Physics,\\
Moscow State University,\\
Moscow 119899, Russia}
\date{}

\begin{document}
\def\pp{\partial_{z}}
\def\ppm{\partial_{\bar{z}}}
\def\pd{\partial_{\mu}}
\def\pdn{\partial_{\nu}}
\def\pu{\partial^{\mu}}
\def\sh{\sinh}
\def\ch{\cosh}
\def\intt{\int\limits_{\Sigma}d^{2}z Tr}
\def\innt{\int\limits_{\Sigma}d^{2}z}
\def\tg{\tilde{g}}
\def\b{\beta}
\def\a{\alpha}
\def\Qp{Q_{+}}
\def\Qm{Q_{-}}
\def\S{\Sigma}
\def\T{1+t^{2}}
\def\te{\theta}
\def\D{\Delta}
\def\l{\Lambda}
\def\Ap{A_{z}}
\def\Am{A_{\bar{z}}}
\def\Jp{J_{z}}
\def\Jm{J_{\bar{z}}}
\def\Ip{I_{z}}
\def\Im{I_{\bar{z}}}
\maketitle

\begin{abstract}
We show how different modifications of Kantowski-Sachs cosmologies emerge
in four dimensions as dimensional reduction of the gauged Wess-Zumino-Witten
model based on  $SO(2,2)/SO(2)$.
\end{abstract}

\vfill
DTP-MSU/97-08\\
hep-th/9704146
\pagebreak

   Nontrivial gravitational backgrounds that can be obtained using gauged WZW
models are of special interest mainly because from them one can extract
information using both conformal field theory techniques and lagrangian methods
\cite{tsey1} .  As conformal field theories, these backgrounds can be taken as
exact solutions of the bosonic string or superstrings\footnote{In the
latter case the Kazama-Susuki construction or heterotic
coset models \cite{hetco} must be used}. The classification of cosets which
have one time coordinate and dimension up to ten has been presented in
\cite{gq}, still there is almost no information on the details of the
theories. In this letter we concentrate on one of the five dimensional cosets
$SO(2,2)/SO(2)$ which upon dimensional reduction to four dimensions yields
different modifications of Kantowski-Sachs cosmologies for early time and for
large values of time gives static universes.

We concentrate on the $SO(2,2)/SO(2)$ coset as it is defined in the axial
gauging since  the vector gauging can be obtained
performing a duality transformation \cite{kir}. The axial gauging of the WZW
model has the following action

\begin{eqnarray}
I(g,A)&=&kI(g)-kI_A \nonumber \\
I(g)&=&\frac{1}{4\pi}\intt \pp g\ppm g^{-1}\nonumber \\
&+&\frac{1}{6\pi} \epsilon^{ijk}\int\limits_{B,\partial B=\Sigma}
d^{3}y g^{-1}\partial_{i}gg^{-1}\partial_{j}gg^{-1}\partial_{k}g \nonumber \\
I_A&=&\frac{1}{2\pi}\intt [\Am g^{-1}\pp g+\Ap\ppm gg^{-1}\nonumber \\
&+&\Ap^{R}g\Am^{L}g^{-1}+\Ap\Am].
\end{eqnarray}
The parametrization of the group element $g \in SO(2,2)$ is as in
\cite{bars}, i. e.,  $g=h\tau$, where $h\in SO(2,1)$ and $ \tau
\in SO(2,2)/SO(2,1)$.  The matrix $\tau$ has the form:

\begin{equation}
\tau=\left(
\begin{array}{cc}
b&(b+1)\hat{X^{\nu}}\\
-(b+1)\hat{X_{\mu}}& \eta_{\mu}^{\nu}-(b+1)\hat{X_{\mu}}\hat{X^{\nu}}
\end{array}
\right),
\end{equation}
here $\hat{X^{\mu}}$ is the arrow $(x^{1},-x^{2},-x^{3})$,
$b=(1-x^{2})/(1+x^{2})$ and the indices are
contracted with the Minkowski metric $\eta_{\mu\nu}=diag(1,-1,-1)$. The
matrix $h$ is constructed using the Euler parametrization,
$h=g_{34}(\phi_{L})g_{23}(r)g_{34}(\phi_{R})$, here $g_{ij}$ is the
one-parametric subgroup of (pseudo)rotation in the $(i,j)$ plane.
We consider the gauge transformation generated by
$T=L_{34}$; a gauge fixing condition is $\phi_{L}=0$, since the
gauge transformation acts on this variable as $\phi_{L}\to\phi_{L}+\alpha$
where $\alpha$ is the gauge parameter. A convenient change of coordinates is:

\begin{equation}
x^{1}=t\cosh\te, \quad x^{2}=t\sinh\te \cos\psi, \quad x^{3}=t\sinh\te\sin\psi
\end{equation}

In these new coordinates we obtain

\begin{eqnarray}
I(g)&=&\frac{1}{2\pi}\innt\left( \frac{4}{(\T)^{2}}\pp t\ppm t
-\pp r\ppm r +\pp\phi\ppm\phi \right. \nonumber \\
&-&\frac{4t^{2}}{\T}\left(\pp\te\ppm\te
+\sh^{2}\te\pp\psi\ppm\psi \right. \nonumber \\
&-&\cos(\psi-\phi)\pp\te\ppm r +\frac{\sinh
2\te}{2}\sin(\psi-\phi)\pp\psi\ppm r \nonumber \\
&+&2\left. \left.\sinh^2\te\pp\psi\ppm\phi \right)\right),
\end{eqnarray}
where $\phi=\phi_{R}$.
The currents of the model, which are defined as $\Jm =Tr T \ppm g g^{-1}$ and
$\Jp=Tr T g^{-1}\pp g$  are:
\begin{eqnarray}
\Jp&=&-\frac{2t^2}{\T}\sh 2\te\sin(\psi-\phi)\pp r -\frac{2(1+t^{2}\cosh
2\te)}{\T}\pp\phi +\frac{4t^{2}\sinh^2\te}{\T}\pp\psi \nonumber\\
\Jm&=&-\frac{2t^2}{\T}(\cos(\psi-\phi)\sh r\sh 2\te+2\ch r\sh^2\te)
\ppm\psi \nonumber \\
&-&2\cosh r\ppm\phi-\frac{4t^2}{\T}\sh r\sin(\psi-\phi)\ppm\te.
\end{eqnarray}
another ingredient is  $N_{a}=Tr(T^2+TgTg^{-1})$,
\begin{equation}
N_{a}=-\frac{2t^2}{\T}(\cos(\psi-\phi)\sh r\sh 2\te+\ch r\ch 2\te+1)
-\frac{4\ch^2(r/2)}{\T}.
\end{equation}
The low energy limit can be reached substituting the gauge field
by its equation of motion and shifting the dilaton field by a certain amount,
\begin{equation}
I=I(g)+\innt\frac{\Jp\Jm}{2\pi N_{a}} \quad \Phi=\Phi_0 +\ln N_{a}
\end{equation}
For $t=0$ which means that we move from $SO(2,2)$ to $SO(2,1)$ (see that for
$t=0$ we have  $\tau=1$ eq. 2) and after changing the level ($k\to-k$) we
reproduce the result of \cite{witb} where $SO(2,1)/SO(1,1)$ was
considered.  Although the five dimensional background that emerges as the low
energy limit of this construction is interesting  in itself we concentrate on
the following situation. As can be seen from eqs. (4-6) the background is
independent of one variable, giving us the possibility of dimensionally
reducing the background to four spacetime dimensions. In this letter we are
going to focus on the emerging four dimensional backgrounds.

The following change of coordinates
\begin{equation}
\psi-\phi\to\psi  \qquad \phi\to\kappa
\end{equation}
makes the fact that the 5d background is independent of $\kappa$
explicit. The dimensional reduction can be made following
\cite{sen} and the resulting metric for small values of $t$ is

\begin{eqnarray}
ds^{2}&=&dt^{2}-\frac{1}{4}dr^{2}-t^{2}(d\te^{2}+\sh^{2}\te d\psi^{2}) \nonumber \\
&-&t^{2}(\cos\psi d\te d r+\frac{\sh 2\te}{2}\sin\psi d\psi d r).
\end{eqnarray}
This metric can locally be represented as a Kantowski-Sachs model if we
change:  $r\to 2r, d\te \to d\te+\cos\psi dr, d\psi \to d\psi
+\coth\te\sin\psi dr$. A similar metric, namely $\psi\to \psi-\pi/2$,
appeared in a recent study of gauge WZW models based on $SO(2,2)/SO(2)\times
SO(1,1)$ \cite{pando} and was shown to describe a modified Kantowski-Sachs
anisotropically expanding universe. This is a very interesting case since in
general dimensional reductions of WZW models do not coincide with gauging the
corresponding symmetry, an example is given in \cite{john1}. In the case
under discussion the relation is obtained because both contributions to the
metric (from dimensional reduction and for gauging the symmetry) are of the
order $O(t^4)$.  The rest of the background includes $U(1)$ gauge fields
according to $A_{\mu}=\frac{1}{2}g_{\mu\kappa}g^{\kappa\kappa}$
\begin{eqnarray}
A_r&=&2t^2\sin\psi\coth(r/2) \nonumber \\
A_m&=&-t^2\sh 2\te\sin\psi/2\sh^2(r/2) \nonumber \\
A_\eta&=&t^2\sh 2\te\cos\psi\coth(r/2),
\end{eqnarray}
and an antisymmetric tensor
\begin{eqnarray}
B_{r\te}&=&-4t^2\cos\psi \nonumber \\
B_{r\psi}&=&2t^2\sh 2\te\sin\psi  \nonumber \\
B_{\te\psi}&=&0
\end{eqnarray}
The dilaton field in the small $t$ limit is $\Phi=2\ln\sh(r/2)$.
For large values of $t$ the theory describes a static universe, this can be
seen taking the limit of large $t$ in (4-6) and changing $t\to 1/t$.  If
instead of the change (8) we make

\begin{equation}
\psi-\phi\to\psi  \qquad \psi\to\kappa,
\end{equation}
in the small $t$ limit for the metric we find
\begin{eqnarray}
ds^2&=&dt^2-\frac{1}{4}dr^2-t^2(d\te^2+\frac{1}{2}\sh^2\te d\psi^2) \nonumber\\
&-&t^2\left(\cos\psi d\te dr-\frac{\sh 2\te\sin\psi}{4}
(\tanh^2(r/2)+2)drd\psi\right.\nonumber\\
&+&\left. \frac{1}{4}(\ch 2\te\tanh^2(r/2)+1)d\psi^2\right)
\end{eqnarray}
This is another modification of a Kantowski-Sachs metric with negative
curvature. Since $\sqrt{-G}=4t^2\sqrt{\ch 2\te(1+\tanh^2(r/2))}$ we see that
the metrics describes an expanding universe. The anisotropy is shown by the
fact that a distance measured only in the $r$ direction does not depend on
time while for any other direction $\te$, for instance, we have $t\Delta\te$.
The componets of the $U(1)$ gauge field are
\begin{eqnarray}
A_r&=&-t^2\sh 2\te\sin\psi(1+coth^2(r/2) \nonumber \\
A_\te&=&2t^2\sin\psi\coth(r/2)\nonumber \\
A_\psi&=&-1/2,
\end{eqnarray}
the antisymmetric tensor componets are
\begin{eqnarray}
B_{r\te}&=&-4t^2\cos\psi \nonumber \\
B_{r\psi}&=&t^2\sh 2\te\sin\psi(1+\tanh^2(r/2)) \nonumber\\
B_{\te\psi}&=&4t^2\sin\psi\tanh(r/2).
\end{eqnarray}
It can be the case that some of the backgrounds presented here are not
conformally invariant. We remind that we have presented some limits (small
and large values of time) of conformally invariant theories which need not to
be conformally invariant themself\footnote{The typical example is the WZW
model on a group manifold and the same model restricted to a homogenous space
obtained from the given group}. To restore conformal invariance at the one
loop level (large $k$ in eq. 1) it might be necessary to account for all the
powers of $t$.  Nevertheless from the geometrical point of view, the
backgrounds we have presented here completely determine the physics of
spacetime at the given limits.

\begin{center}
\large Acknowledgments
\end{center}

I am very grateful to D. V. Galtsov and M.Z. Iofa for many useful
discussions.

\pagebreak

\end{document}